\documentclass[a4paper,10pt]{article}
\usepackage{theorem,amssymb}
\usepackage{graphicx}
\def\bE{{\mathbf E}}
\def\Var{\mathrm{Var}\,}

\def\<{\langle}
\def\>{\rangle}

\begin{document}
\title{Light-matter quantum interferometry with homodyne detection}

\author{L\'aszl\'o Ruppert$^{1,*}$ and Radim Filip$^{1}$\\
\small $^{1}$Department of Optics, Palacky University, 17. listopadu 12, 771 46 Olomouc, Czech
Republic\\
\small\textit{$^{*}$ruppert@optics.upol.cz}} 
\date{}

\maketitle

\begin{abstract}
We investigated the estimation of an unknown Gaussian process (containing displacement, squeezing and phase-shift) applied to a matter system. The state of the matter system is not directly measured, instead, we measure an optical mode which interacts with the system. We propose an interferometric setup exploiting a beam-splitter-type of light-matter interaction with homodyne detectors and two methods of estimation. We demonstrate the superiority of the interferometric setup over alternative non-interferometric schemes. Importantly, we show that even limited coupling strength and a noisy matter system are sufficient for very good estimation. Our work opens the way to many future investigations of light-matter interferometry for experimental platforms in quantum metrology of matter systems.
\end{abstract}




\section{Introduction}

The rapid development of quantum technology crucially depends on the quality of quantum metrology. Due to recent progress in the field, optical interferometry using light as a probe modified in the optical phase by a matter sample \cite{int1,int2} can be combined with quantum interfaces and transducers between matter and light or microwaves \cite{rmp1,rmp4,rmp2,rmp3}. A steady experimental development of quantum interfaces between light and atomic ensembles  \cite{inta1,inta2,inta3,inta4,inta5,inta6,inta7,inta8,inta9} was recently followed by a dynamic experimental boom of electromechanical and optomechanical interfaces \cite{intm0,intm1,intm2,intm3,intm4,intm5,intm6,intm7}, the extension of which is expected to continue. The combinations of light-matter interfaces and optical interferometry allow a use of matter at quantum level as a sensitive probe of physical processes caused by its environment. Light (or microwaves) only prepares and reads out the quantum state of the matter probe. This is an interesting extension of a broad class of quantum interferometry and metrology \cite{met1,met2,met3}, where the state of light is only changed by unitary dynamics and not at all transferred to matter and passed back on to light. A pioneering example of the extended strategy is quantum magnetometry with atomic ensembles already exploiting nonclassical light and correlations transferred to atomic ensembles \cite{mag1,mag2,mag3,mag4,mag5}. Recently, this development has stimulated first considerations in quantum sensing of mechanical motion and other physical quantities. These approaches broaden the horizons of quantum metrology with correlated and potentially nonclassical matter probes in addition to nonclassical light reading it out.

The basic configuration of light-matter interferometry investigated here has a fully optical path and a matter path. Light-matter quantum interferometry consists of four principal steps: first, a light-matter interface which can partially transfer quantum states of light to typically thermal states of matter; second, the quantum state of matter is modified by the process coming from the environment (the estimation of this process is the objective of our work); third, a light-matter interface which can partially transfer the modified matter state back to light; fourth, the quantum state of the light is measured and the matter process is estimated. Such an arrangement allows to discuss and compare various basic configurations of light-matter interferometry, depicted in Figs.~\ref{scheme} and \ref{schemeB}. In contrast to standard optical interferometry, two experimental limitations are commonly present. Light is typically coupled with matter only weakly, the complete ideal swap of the quantum state of the light to matter and back is limited by decoherence, damping and noise. These set a limit on the interaction strength of the light-matter interface if it should be close to unitary quantum coupling. Moreover, the matter state is typically noisy and cooling it to the ground state can be limited, especially in the early stage of experiments. In addition, only one output of the light-matter interferometry is measured, because the matter state cannot be measured well without an additional noise due to the weakness of light-matter coupling. Such limitations are not present in optical interferometry \cite{int3} due to the almost ideal performance of the beam splitters (BS), the high mode matching, the vacuum states in the empty port of the beam splitter and direct access to both light beams after the interferometer by intensity or homodyne detectors. The experimental limitations of the light-matter interface therefore raise many questions, mainly, if it is possible to perform an estimation of the matter process for an arbitrarily weak coupling with matter noise. The answer to the main question about the quality of realistic light-matter quantum interferometry is an important step before specifics of experimental platforms can be analyzed.

We therefore propose and analyze the estimation of the processes changing a matter oscillator (mechanical, electrical, atomic) by an environment using a realistic light-matter interferometry with coherent light and homodyne detection. We use a beam splitter type of light-matter interaction common for experimental platforms with matter oscillators \cite{rmp1,rmp2} and analysis of optical interferometry \cite{int1,int2,int3}. We focus on the estimation of a general Gaussian process \cite{GP1,GP2,GP3,GP4,GP5} in the phase space of a single matter oscillator (includes simultaneously a phase-shift, a displacement and a squeezing of the oscillating amplitudes). It covers a broad class of weak physical couplings of a matter probe to an environment which can be linearized in the amplitudes of oscillations. Different estimators are investigated for various setups and their efficiencies are compared for different parameters. Our results show that light-matter interferometry is in principle possible and therefore they open the way for more technical investigations for specific experimental platforms.

\section{Preliminaries}

\subsection{The model}

\begin{figure}[!t]
\begin{center}
\includegraphics[width=0.7\columnwidth]{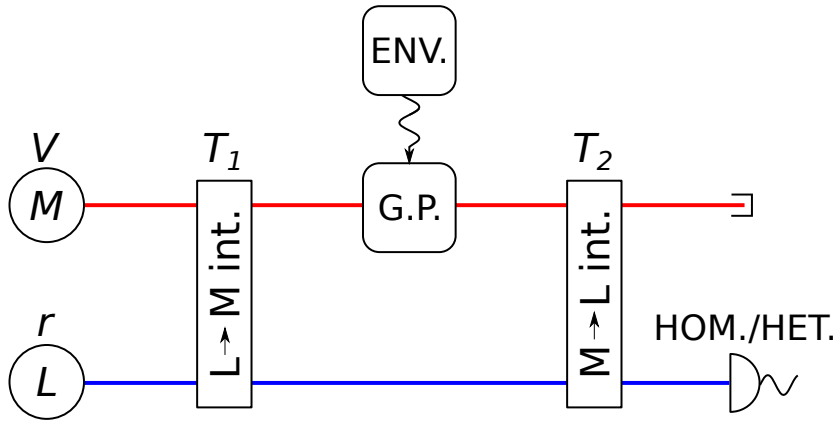}
\caption{\label{scheme} Schematic representation of light-matter interferometry with quantum interfaces. A material sample (mode $M$) in thermal state with variance $V$ interfaces with a coherent light (mode $L$) with amplitude $r$; their interaction is approximated well by a beam-splitter (BS) with transmittance $T_1$. Further on, the matter mode is modified by an unknown Gaussian process (G.P.) stimulated by the environment. The two modes then interact again on a second BS-type interface with transmittance $T_2$. Finally, the quadratures of the light are measured with heterodyne (or homodyne) detection (HOM./HET.) to estimate the parameters of the Gaussian process. The other, matter mode is not measured.}
\end{center}
\end{figure}

The investigated scheme is shown in Fig.\ \ref{scheme}. We have two independent modes, the first is a matter mode $M$ which is in thermal equilibrium with variance $V$, and the second a coherent light mode $L$ with amplitude $r$. Let $a$ and $a^\dagger$ be the creation and annihilation operators for the matter mode $M$, $b$ and $b^\dagger$ the same for the light mode $L$. We use an interferometric type of setting (we will later compare this with other scenarios as well), meaning that we couple the two modes, apply the unknown Gaussian process to the matter state, and use a coupling again. Finally, in contrast to classical interferometry, we cannot access both modes, we can only measure the light mode. If we use a homodyne or heterodyne measurement (on $N$ independent states) we can directly access the quadrature values of the light.

In general, light-matter interaction can be described by a linearized process. However, if we are in the sideband-resolved regime (where the mechanical frequency is much higher than the cavity decay rate), we have three interesting special cases, one of which is the so-called red-detuned regime \cite{rmp4,intm0}. In this case we have two harmonic oscillators of nearly equal frequency, and the linearized Hamiltonian will have a leading term which corresponds to an optical beam splitter (BS). This BS interface has been already obtained between radiation and atoms \cite{rmp4} as well as mechanical oscillators, and is actually ubiquitous in optomechanical experiments \cite{intm0}. These interactions are typically weak, so the transmittance of the BS interface should be considered low ($T_1, T_2 \ll 1$). 

We decompose the Gaussian unitary process induced by the environment on the matter mode $M$ into three parts:
\begin{equation}\label{channel_form}
\rho^*=D(\gamma)R(\Phi)S(\xi)\rho S(\xi)^\dagger R(\Phi)^\dagger D(\gamma)^\dagger,
\end{equation}
where  $S(\xi)=\exp(1/2 \xi^2 a^{\dagger2}-1/2 \xi^{*2} a^2)$ is the squeezing operator (with $\xi=w \mathrm{e}^{i \alpha}$), $R(\Phi)=\exp(i \Phi a^{\dagger}a)$ is the phase-shift operator, and $D(\gamma)=\exp(\gamma a^{\dagger}-\gamma^* a)$ is the displacement operator (with $\gamma=d \mathrm{e}^{i \beta}$). The squeezing is described by the magnitude $(q=\mathrm{e}^w)$ and by the direction $(\alpha)$ of the squeezing. The displacement is described by its magnitude $(d)$ and its direction $(\beta)$. The combination of these three transformations describes each possible Gaussian transformation of a given Gaussian state which preserves the mixedness of the original state. We investigated general Gaussian processes, since they are sufficient approximations for linearized processes in the matter.  

We assume that all parameters of the scheme (i.e., $V, r, T_1, T_2$) are known, however, there are limitations. The effective transmittances $T_1$, from the light mode $L$ to the matter mode $M$, and $T_2$, from matter mode $M$ to light mode $L$ are limited to small values by decoherence and thermalization. At the same time, the variance $V$ of matter noise is typically large at the room temperature. The amplitude $r$ of coherent light can be increased if necessary. Our task is to estimate the parameters of the unknown Gaussian process, namely, $\Phi, q, \alpha, d, \beta$.

\subsection{Quantifying estimator performance}

Before investigating dependence of the performance of different estimators on other parameters of the scheme, a figure of merit must be chosen. This choice is, however, not self-evident. The use of the popular quantum Fisher information is undesirable in the current scenario since it provides us only an upper bound of the obtainable information, which could suggest the optimality of experimentally not feasible or simply impractical measurements. Thus, since we are interested in using simple optical homodyne measurements, it makes more sense to incorporate also this measurement setup into the calculation, that is, calculate the classical Fisher information or the variance of the estimator instead.

The Fisher information for parameter $\theta$ with a given likelihood function $f$ can be calculated as \cite{CR}
\begin{equation}
I(\theta)=\bE_{\underline{x}}\bigg(\frac{\partial}{\partial\theta} \log f(\underline{x};\theta)\bigg)^2=-\bE_{\underline{x}}\bigg(\frac{\partial^2}{\partial\theta^2} \log f(\underline{x};\theta)\bigg).
\end{equation}

In general, a larger Fisher information means more information about the given parameter. However, it is independent of the used estimator, it provides us instead a theoretical bound on the variance by the Cram\'er-Rao inequality \cite{CR}:
\begin{equation}
\Var(\hat\theta)\ge I^{-1}(\theta).
\end{equation}

The variance of a non-linear estimator is difficult to calculate analytically, hence we also used the empirical mean squared error (MSE):
\begin{equation}
MSE_M(\hat{\theta}):=\frac{1}{M}\sum_{k=1}^M(\hat{\theta}_k-\theta)^2,
\end{equation}
where $M$ denotes a technical parameter: the number of repetitions of the numerical simulations (we used $M=10^4$). That is, to obtain the MSE we generated $M$ different realizations of the estimator independently, and checked their average squared distance from the real parameters. Note that the number of measurements $N$ is not related to this quantity, it is the number of measurements in a single realization. The connection to the Cram\'er-Rao bound is that we have in limit $MSE_M(\hat{\theta})\approx MSE(\hat{\theta})\approx \Var(\hat{\theta})$, where the first approximation is true if $M$ is large (law of large numbers), while the second approximation is true if $N$ is large (the estimator $\hat{\theta}$ is asymptotically unbiased).

As we are working with the first and second moments of Gaussian states the only problem that can occur is that  due to the measurement errors the estimated covariance matrix will correspond to a non-physical state, that is, it will not fulfill the Heisenberg uncertainty relation \cite{QI}. This can basically occur only if the thermal state of matter is close to vacuum and the number of measurements ($N$) is low. That is, in a practical setup this is unlikely to happen, but even if it does, we can map an unphysical covariance matrix to a close, physical one.

\section{Estimation of the displacement}

Before we handle the general case, let us investigate some special cases first. The most simple, but still relevant case is when the unknown Gaussian process consists only of a displacement. In terms of the parameters, this means that we can assume that $q=1$, $\Phi=0$, and the only parameters of interest are $d$ and $\beta$, the magnitude and direction of the displacement vector. Note that this is an important scenario, since it corresponds to the short-time impact of a constant external force. 

\begin{figure}[!t]
\begin{center}
\includegraphics[width=0.45\columnwidth]{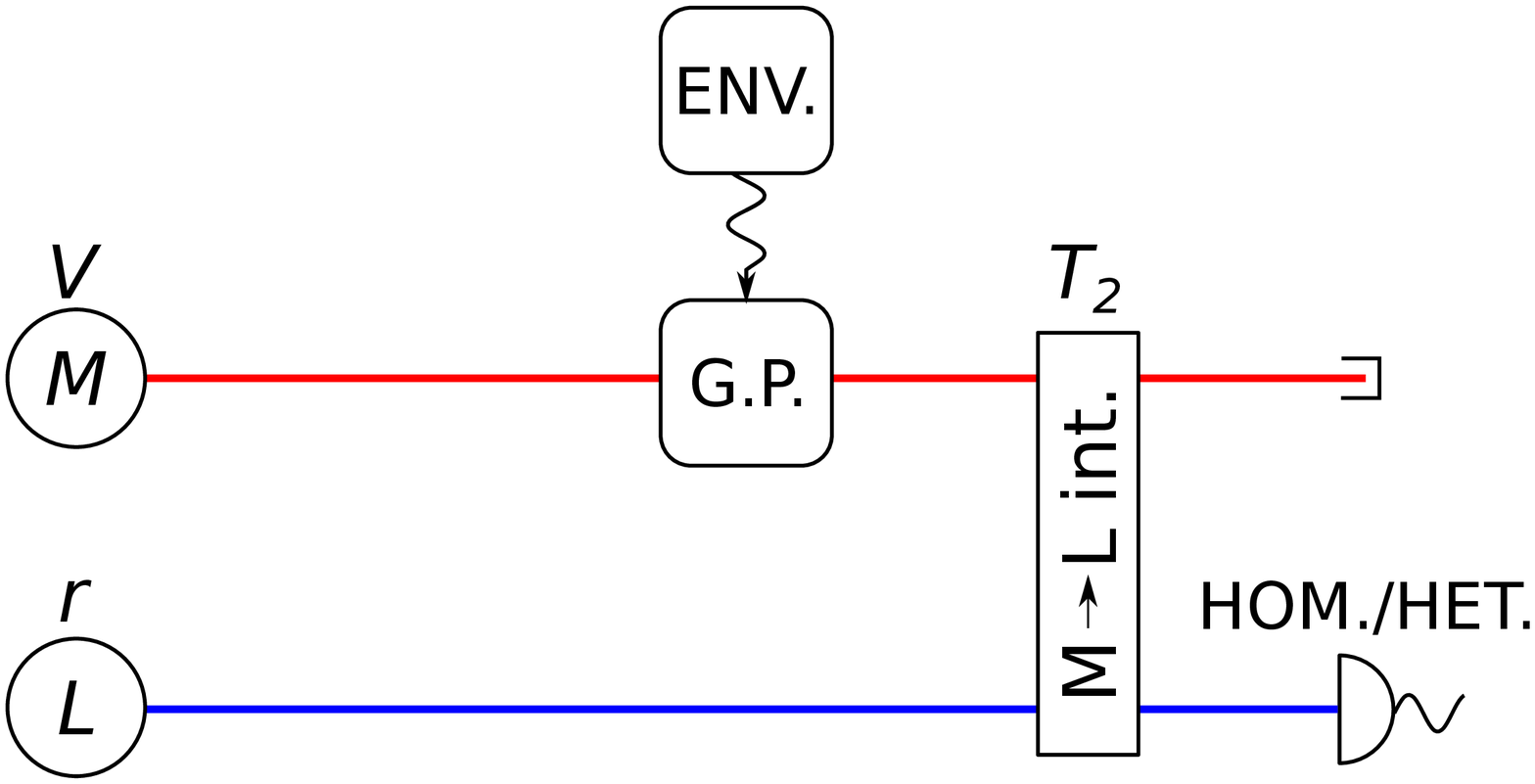}
\includegraphics[width=0.45\columnwidth]{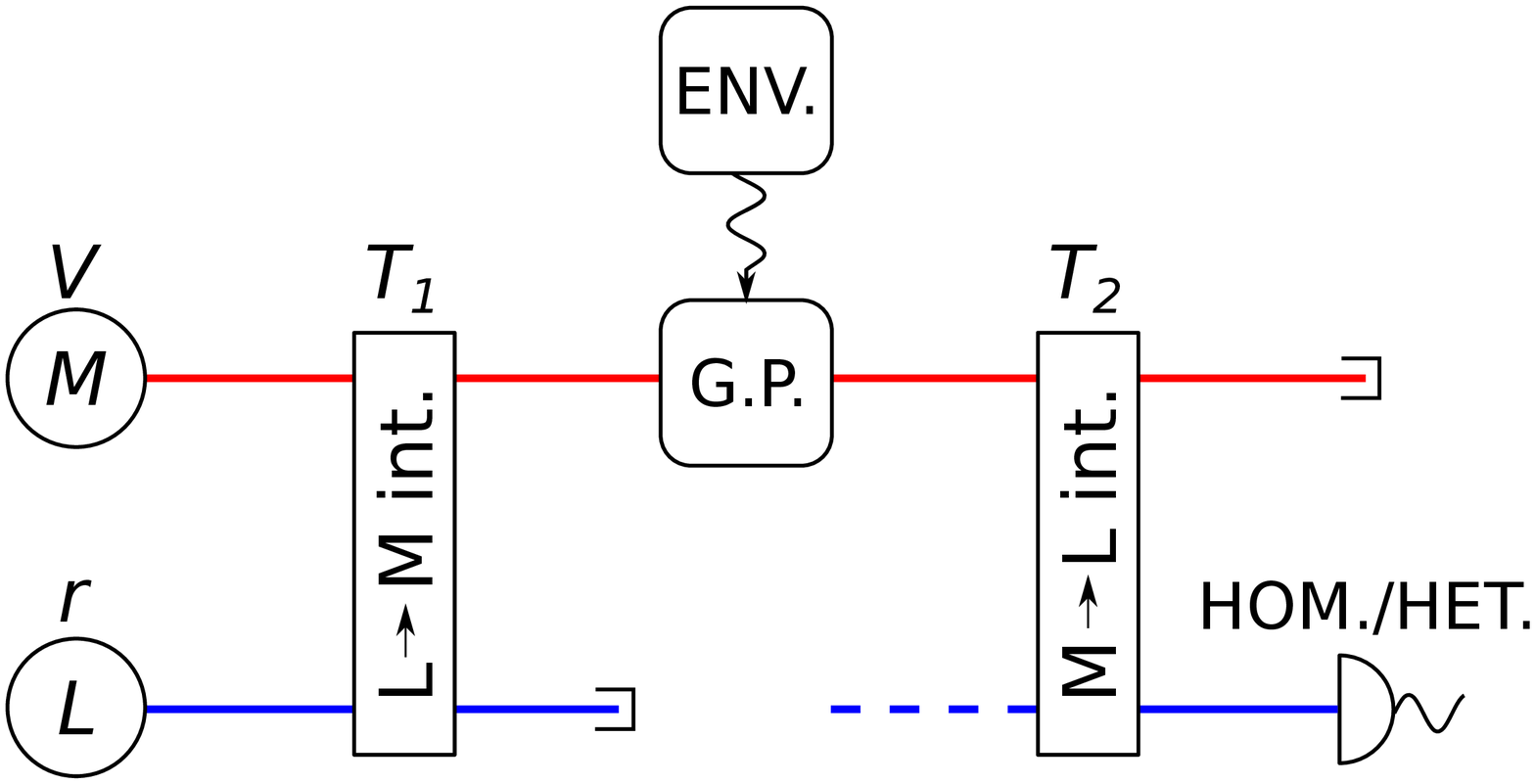}
\caption{\label{schemeB} Variations of non-interferometric setups: (left) the most simplistic setup, where the the interface is applied only after the measured unknown Gaussian process, (right) light interfaces with the matter at the first BS interaction, but is blocked after that, so the coherent light is only used for the preparation of the state, not for read-out by the second interface. (We used the notations of Fig.\ \ref{scheme}.)}
\end{center}
\end{figure}

We can perform the estimation of these parameters even in the most simplistic scenario (Fig.\ \ref{schemeB}, left subfigure): the displacement is applied directly to the matter mode, we couple this with the optical mode, and from the change of the optical mode we can estimate the desired parameters. And we can use an alternative version (Fig.\ \ref{schemeB}, right subfigure): the light can be coupled with the matter on the first BS, but after that it is blocked, so on the second BS the optical mode is in vacuum state. For the sake of complete analysis of non-interferometric schemes we can mention a fourth case, in which we use a coherent state neither for preparation nor for measurement (i.e., having the simplistic scenario but using a vacuum source for light), but this is equivalent to the simplistic scenario shown in Fig.\ \ref{schemeB}, left subfigure.

For a more accurate description let us denote the $x$-quadrature ($(b+b^\dagger)/\sqrt{2}$) of the mechanical mode with $x_1$, the $x$-quadrature of the coherent light with $x_2$. Let us assume that the phase of the coherent state is in the $x$-direction. Then $\<x_1\>=0$, $\Var(x_1)=V$, $\<x_2\>=r$, $\Var(x_2)=1$. 

In the simplistic scenario the unknown process changes the mechanical mode directly: $x_1^*=x_1+d_x$, where $d_x=d\cos\beta$. The measured mode after the beam splitting will be
\begin{equation}
x_o=\sqrt{T_2} \cdot x_1^*+\sqrt{1-T_2} \cdot x_2.
\end{equation}

If we take the mean of this equation, we get $\<x_o\>=\sqrt{T_2} \cdot d_x+\sqrt{1-T_2} \cdot r$, from which we can simply estimate the unknown displacement:
\begin{equation}
\widehat{d_x}=\frac{\<x_o\>-\sqrt{1-T_2} \cdot r}{\sqrt{T_2}}.
\end{equation}
Similarly, for the $p$-quadratures ($\<p_2\>=0$ and $d_p=d\sin\beta$) we obtain $\<p_o\>=\sqrt{T_2} \cdot d_p$ and obtain  the estimator
\begin{equation}
\widehat{d_p}=\frac{\<p_o\>}{\sqrt{T_2}}.
\end{equation}
The estimator of $d$ and $\beta$ can be easily constructed using the previous estimators: $\hat d=\sqrt{\widehat{d_x}^2+\widehat{d_p}^2}$ and $\hat \beta=\arctan(\widehat{d_p}/\widehat{d_x})$.

Let us calculate the Fisher information for parameter $d$ (the conclusions for $\beta$ are very similar). The likelihood function is given on a two dimensional plane of quadratures: $\underline{x}=(x,p)$ and for large $N$ it can be approximated well (due to the central limit theorem) with a Gaussian distribution, the first moments of which are already calculated, while its second moments are $\Var(x_o)=\Var(p_o)=T_2\cdot V+(1-T_2)$.
In this particular situation the second derivative of $\log(f)$ will be constant, so we can easily obtain the Fisher information:
\begin{equation}\label{one}
I(d)=\frac{T_2}{1+T_2(V-1)}.
\end{equation}
It is also easy to check that we have $\Var(\hat{d})=I^{-1}(d)$, so the Cram\'er-Rao bound is saturated, the obtained estimator is efficient. We can see that the estimation is more precise, i.e., Fisher information (\ref{one}) increases, if $T_2$ is larger, or $V$ is smaller (and does not depend on $r$). For a large $V$, $I(d)$ converges to $1/V$, therefore the noise of the matter limits the Fisher information.  

However, the given estimation is only optimal with the given scheme, even better results can be achieved using a different setup: if we apply an additional BS before the unknown process, and block the optical mode somewhere before the second BS (Fig.\ \ref{schemeB}, right subfigure). We can calculate the Fisher information similarly, and obtain 
\begin{equation}\label{two}
I(d)=\frac{T_2}{1-T_2+T_2((1-T_1)V+T_1)}.
\end{equation}
For a given $T_2$ this will be maximal for $T_1=1$ (then having $I(d)=T_2$), but that means we can transfer the state of the optical mode completely to matter, but in practice both $T_1$ and $T_2$ are limited, in which regime the efficiency of this setup is close to the previous one's (Fig.\ \ref{ind}, left subfigure, dashed vs dotted line). For large values of $V$ the Fisher information converges to $1/V/(1-T_1)$, resulting in an increase the magnitude of which depends on the value of $T_1$. 

For the interferometric scheme we can also calculate the values analytically, the minimal value is taken for given $T_2$ if $T_1$ has the same value. Using the notation $T:=T_1=T_2$, we have simply
\begin{equation}
I(d)=T,
\end{equation}
which is the same as in $(\ref{one})$ with $V=1$, or in $(\ref{two})$ with $T_1=1$. The reason behind this is simple: in the previous cases the thermal variance of the matter passes to the measurement causing a higher uncertainty, while if we use an interferometric scheme, this effect can be eliminated completely. This means that if the variance of the thermal mode is initially high, we can improve the estimation significantly using an interferometric scheme up to a Fisher information only limited by $T$ (see Fig.\ \ref{ind}, left subfigure, solid line). It is very stimulating, since we can obtain metrology of matter without cooling it down.

\section{Estimation of the phase-shift}

Let us compare it to the case where the unknown Gaussian process contains only a phase-shift. That is, we can assume that $q=1$, $d=0$ and the only parameter we are interested in is $\Phi$.

In this case the simplistic scheme is not useful at all, since if we apply a phase-shift to a thermal state and then couple it with a coherent beam, then the measurement outcome will be the same, independently from the phase-shift parameter $\Phi$.

\begin{figure}[!t]
\begin{center}
\includegraphics[width=0.45\columnwidth]{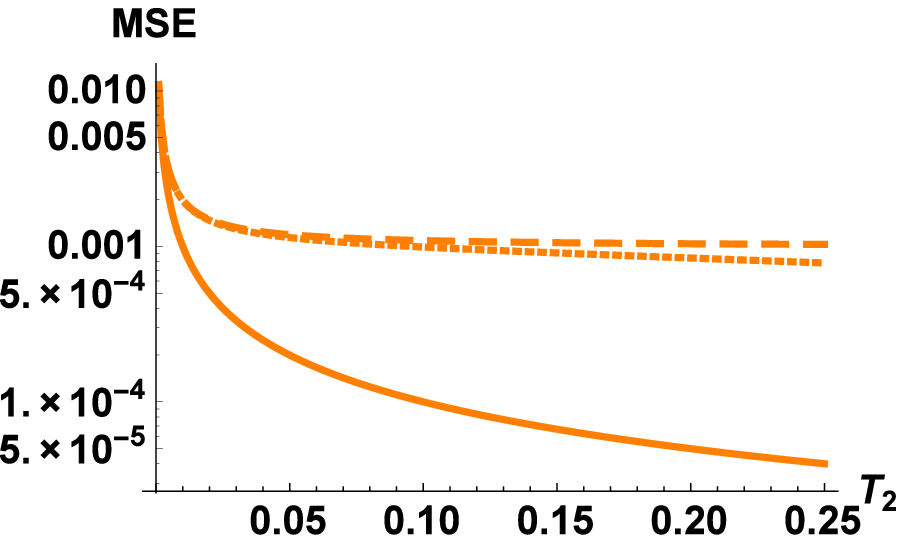}
\includegraphics[width=0.45\columnwidth]{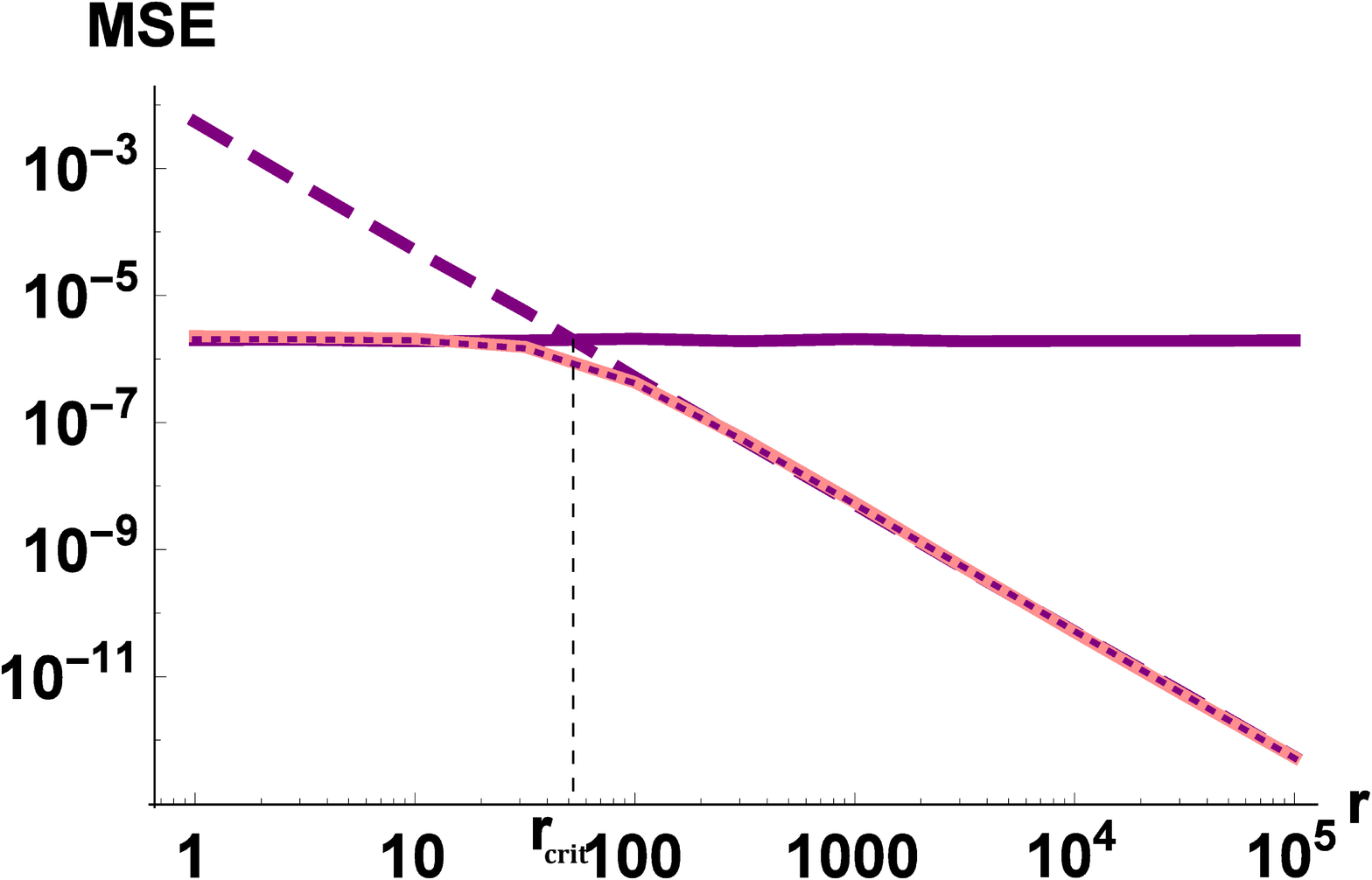}
\caption{\label{ind} (left) The MSE of the estimators of the displacement ($d$) if only displacement is present. The solid line corresponds to interferometric setup with $T_1=T_2$, the dashed line corresponds to simplistic setup ($T_1=0$), the dotted line corresponds to the case of blocked optical beam with $T_1=T_2$. We have used the following parameters: $V=100$, $N=10^5$, $d=4$, $\beta=0.5$. (right) The MSE of the estimators of the phase-shift ($\Phi$) if only phase-shift is present. The solid line comes from the estimators using only the covariance matrix (second moments), the dashed line comes from the estimators using only the means (first moments),  the dotted line comes from the maximum likelihood estimator, the red (lighter) solid line is the theoretical limit coming from Cram\'er-Rao inequality (note the last two lines basically coincide). We have applied the following parameters for the scheme: $T_1=T_2=0.1$, $V=100$, $N=10^5$, $\Phi=0.7$.}
\end{center}
\end{figure}

On the other hand, using the interferometric scheme we can estimate $\Phi$ in two different ways, since the phase-shift changes both the first and the second moments of the measured mode. The variance of the output mode is
\begin{equation}
\Var(x_o)=\Var(p_o)=u+v\cos(\Phi),
\end{equation}
where $u=1-T_1-T_2+2T_1T_2+T_1V+T_2V-2T_1T_2V$ and $v=2(1-V)\sqrt{(1-T_1)(1-T_2)T_1 T_2}$.
So we obtain the estimator
\begin{equation}
\hat\Phi_1=\arccos\Bigg(v^{-1}\bigg(\frac{\Var(x_o)+\Var(p_o)}{2}-u\bigg)\Bigg)
\end{equation}

The other alternative is using the means of the output mode, which are $\<x_o\>=r\sqrt{(1-T_1)(1-T_2)}+r\sqrt{T_1T_2}\cos{\Phi}$ and $\<p_o\>=r\sqrt{T_1T_2}\sin{\Phi}$. From these we obtain the estimator
\begin{equation}
\hat\Phi_2=\arctan\left(\frac{\<p_o\>}{\<x_o\>-r\sqrt{(1-T_1)(1-T_2)}}\right).
\end{equation}

We can also calculate the Fisher information by using a Gaussian approximation of the output, but in this case it can be done only numerically. We can see in the right subfigure of Fig.\ \ref{ind} that for small values of $r$ it is the first estimation method ($\hat\Phi_1$), while for the large values of $r$ it is the second estimator ($\hat\Phi_2$) that performs close to the theoretical optimum. It is not surprising that the two discussed methods have different optimal working regimes, since using the means for estimation purposes is only preferable if their values are large enough to detect a change in them. The change of the variance is, in contrast, preferable if the thermal source is stronger than the coherent one.

The first estimator does not depend on the parameter $r$, while the second estimator has a MSE which is proportional to $1/r^2$. So there is a critical value of $r$ (denote it with $r_{\mathrm{crit}}$), where the MSEs of two methods are the same, the first is better if $r<r_{\mathrm{crit}}$ and the second estimator is better if $r>r_{\mathrm{crit}}$. Unfortunately, it is hard to tell in general the actual value of $r_{\mathrm{crit}}$, since it depends on the other parameters too. Nevertheless, it is true that for larger values of $V$ the value of $r_{\mathrm{crit}}$ will also increase, while for larger values of transmittances the value of $r_{\mathrm{crit}}$ will decrease. So, in general, it is not trivial to tell, which estimator is better, but if the strength of the coherent light is increased, then using the means for estimation will eventually be better.

We can circumvent numerically the issue of uncertain optimal working regimes by using a maximum likelihood (ML) estimator. This estimator would produce a similar magnitude of error as the theoretical optimum for any value of $r$ (in Fig.\ \ref{ind}, right subfigure, the purple dotted and the red solid lines basically coincide). This is not surprising since the ML estimator provides an asymptotically efficient estimator. Our obtained estimators have different advantages: they are easily obtainable analytically and show a clear mathematical structure of the important quantities in different parameter regions, which helped us obtain practically efficient estimators for every case except a short intermediate interval around $r_{\mathrm{crit}}$.

We can compare this also with the non-interferometric case of the blocked optical beam, and interestingly in this case the situation is not so clear as it was for the displacement: the interferometric scheme is only better if $\cos(\Phi)>0.5$. The reason is simple again, if we change the phase of state too much, the interferometry does not decrease the variance but rather increases, so blocking the optical mode can be actually helpful. The difference is, however, not significant for practical ranges of the scheme parameters (i.e., when $T$ and $V$ are small: $T\ll 1$ and $V\approx 1$). 

We omit the discussion of the case when the Gaussian process only contains squeezing, because it is very similar to the phase-shift only case discussed here and these examples have been presented mainly to demonstrate peculiar features of the light-matter interferometry.  

\section{Estimation of a general Gaussian process}

In what follows, we investigate the estimation of a general Gaussian process simultaneously consisting of displacement, phase-shift and squeezing (with parameters $q,\alpha,d,\beta, \Phi$). Interestingly, we can generalize the two methods used for the estimation of the phase-shift in the previous section, and the obtained estimators will show strong similarities to the already discussed cases. From the previous discussion we can conclude that the interferometric scheme clearly outperforms the simplistic scheme, while blocking the optical beam can be useful in some cases, but it is of minor importance, so in the following we will restrict our investigation only to the interferometric scheme.

\subsection{Estimators}

The displacement of the state only affects the mean of the state, but both phase-shift and squeezing causes a change in the mean and the covariance matrix as well. Using these statistics, one can devise the following methods of estimation:

\textit{Method (i) - estimation using the covariance matrix:} If there is squeezing ($q>1$), then the covariance matrix can be characterized uniquely using the unknown parameters $q, \alpha, \Phi$. Unfortunately, the covariance matrix is a complicated function of the parameters, so the inversion could be done only numerically. Nevertheless, the parameters of the squeezing and the phase-shift can be obtained from the covariance matrix of an arbitrary input state, and then we can check what effect this result has on the mean of the input state. Finally, if we compare the means of the output state and the input state after applying the squeezing and the phase-shift, the difference yields the unknown displacement ($d, \beta$).

\textit{Method (ii) - estimation using only the mean values:} Using a single input we cannot characterize the process simply because of dimensional issues (the mean has 2 parameters, the process has 5 parameters), but  this problem can be overcome if we use multiple input states. For example, if we use two coherent inputs with opposite phases, then the mean after the squeezing and the phase-shift will remain the same with a difference of $\pi$. That is, their sum will be zero, so the average of the output will give exactly the unknown displacement. Similarly, if we apply an input with a phase-shift of $\pi/2$, we can discriminate the effects of the unknown squeezing and the phase-shift by using simple analytical formulas.

\subsection{Dependence on source strengths}

\begin{figure}[!t]
\begin{center}
\includegraphics[width=0.45\columnwidth]{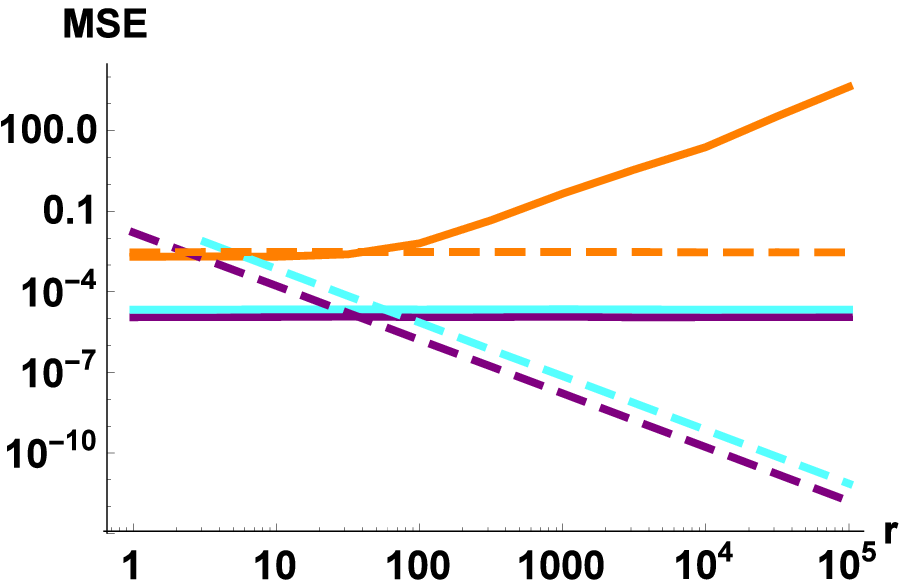}
\includegraphics[width=0.45\columnwidth]{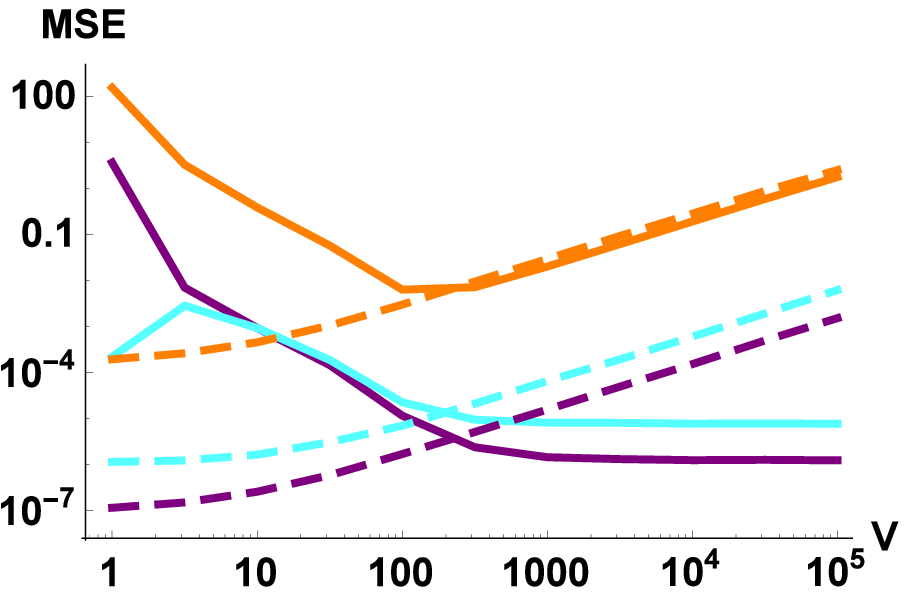}
\caption{\label{rV} The MSE of the estimators of the Gaussian process as a function of (left) the strength of the coherent light ($r$), (right) the variance of the matter ($V$). The purple (dark) lines correspond to the phase shift, the orange (medium) lines correspond to the magnitude of the displacement and cyan (light) lines correspond to the magnitude of the squeezing. The solid lines come from the estimators using the covariance matrix, the dashed lines come from the estimators using only the means. We have applied the following parameters for the scheme: $T_1=T_2=0.1$, $N=10^5$, (left) $V=100$, (right) $r=100$, and for the unknown process: $\Phi=0.7$, $d=4$, $\beta=0.5$, $q=2$, $\alpha=-0.3$.}
\end{center}
\end{figure}

We can see that for Method (i) (Fig.\ \ref{rV} left subfigure, solid lines) increasing the strength of the optical source ($r$) is not helpful. The estimation accuracy of the squeezing and the phase-shift does not change (since the variance of the source does not change either), and a larger displacement of the source means a larger uncertainty in the estimation of the displacement. On the other hand, if the variance of the matter ($V$) is larger the method is quite useful (Fig.\ \ref{rV} right subfigure, solid lines). The estimation of the squeezing and the phase-shift improves when $V$ increases because the input state of the process will be larger and so their effects are better observable (as in standard interferometry). This even compensates the uncertainty of the displacement up to a certain level.

The estimators for Method (ii) behave differently than those derived from the covariance matrix (Fig.\ \ref{rV}, dashed lines). Increasing the strength of our coherent source ($r$) does not have an effect on the estimation of the displacement, since the averaging eliminates the effect of the initial displacement. For the squeezing and the phase-shift we can achieve an improvement because their effect is more visible if the initial amplitude of the coherent light is larger (similarly to the larger variance in the case of the covariance matrix). The larger variance of the thermal source ($V$) in this scenario only adds more noise to our estimation so the errors of all estimators will increase if $V$ is increased.

From the previous analysis we can conclude that the two estimation methods are optimal in different working regimes. A larger $V$ is better for the method based on the covariance matrix, a larger $r$ is better for the method which uses only the mean values. However, the two estimation procedures can be performed simultaneously: when using the second method, the first method can be also applied on all three input states and one can calculate the average of the three sets of estimates.
This has several advantages: first, we can check whether our model is consistent with reality. Since both our estimation methods give us asymptotically unbiased estimators, for large values of $N$ they should be close to the real values. So if there is a large difference in the values coming from the two estimation methods we can be sure that there is some significant effect which has been neglected. Furthermore, we can combine the two estimators in order to decrease estimation variance. The combination can produce high estimation efficiencies for the squeezing and the phase-shift for any value of $V$ (see Fig.\ \ref{rV}, right subfigure). 

In general, it is better if we use a stronger coherent light ($r$ is large) because that can improve the estimation of the squeezing and the phase-shift significantly. If we are interested in the estimation of the displacement (but other parameters are simultaneously unknown) we can conclude that it does not depend on the strength of the coherent light, but a larger variance of the thermal matter results in a less accurate estimation.

\subsection{The effect of coupling strength}

\begin{figure}[!t]
\begin{center}
\includegraphics[width=0.45\columnwidth]{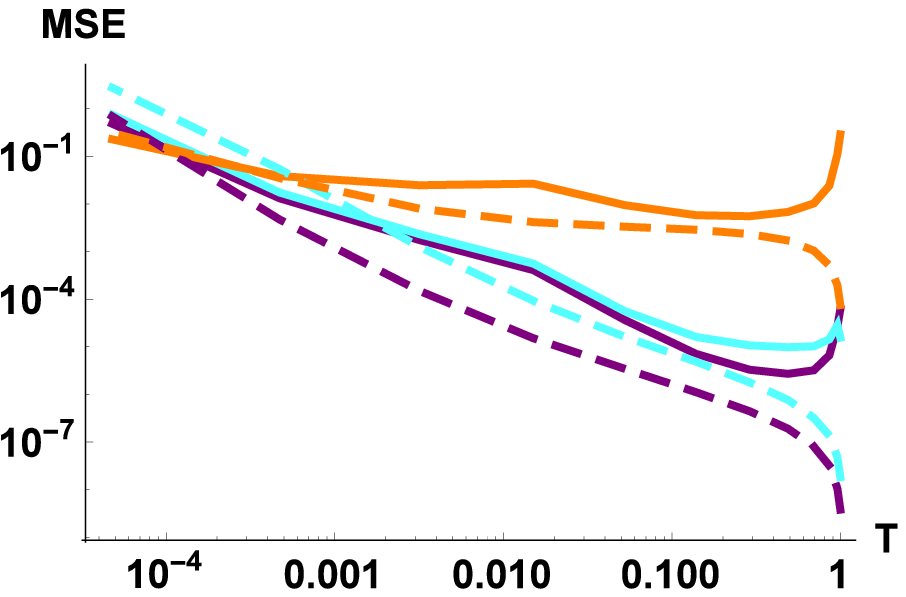}
\includegraphics[width=0.45\columnwidth]{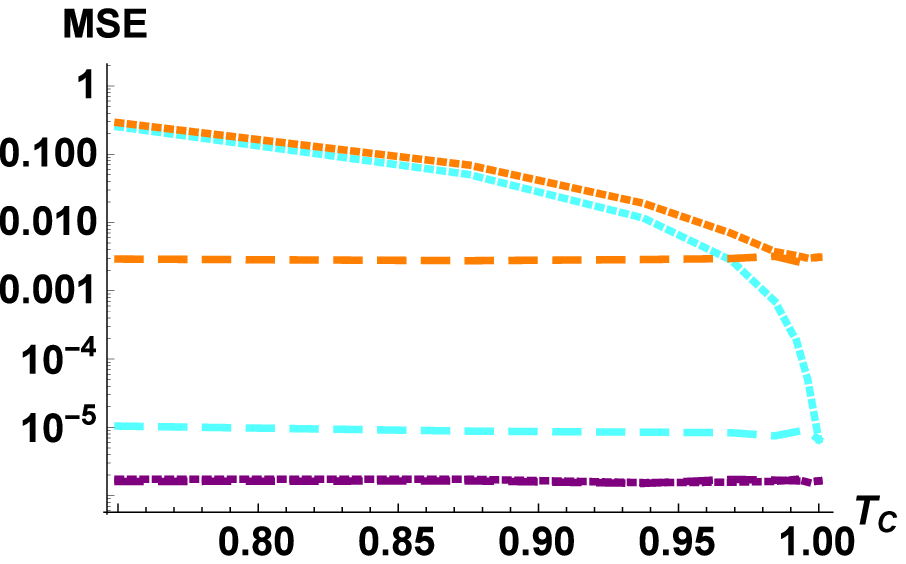}
\caption{\label{T1} The MSE of the estimators of the Gaussian process as a function of (left) $T=T_1=T_2$ and (right) loss. The purple (dark) lines correspond to the phase shift, the orange (medium) lines correspond to the magnitude of the displacement and cyan (light) lines correspond to the magnitude of the squeezing. The solid lines come from the estimators using the covariance matrix, the dashed lines come from the estimators using only the means, the dotted lines come from the naive estimators (assuming that the channel is ideal, i.e. not using a calibration phase) using only the means (note that the purple dashed and the purple dotted lines coincide in the right subfigure). We have applied the following parameters for the scheme: $r=100$, $V=100$, $N=10^5$ (right) $T=0.1$, $V_C=1.2$ and for the unknown process: $\Phi=0.7$, $d=4$, $\beta=0.5$, $q=2$, $\alpha=-0.3$.}
\end{center}
\end{figure}

Moving on from the properties of the initial states to the properties of the optical elements, and let us investigate how the estimation efficiency improves if the coupling is stronger. Let us assume that we have the same effect on either optical element ($T_1=T_2=:T$). We can see (Fig.\ \ref{T1}, left subfigure) that all estimators naturally improve (except for very high values of $T$) if the coupling is stronger. We can even determine the order of the MSE as a function of $T$ around zero, that is, $MSE \sim T^{-c}$: for displacement estimators it is $c\approx$ 0.5-1, for squeezing and phase-shift estimators it is $c\approx$ 1.5-2. Mainly, the estimation performs acceptably well even for small values of $T$ (less than $0.1$) despite $V$ being large.

Let us also note that the optimal performance is achieved for Method (i) around $T=0.5$, that is, when the interferometry is symmetrical. On the other hand, for Method (ii) the optimum is achieved at $T=1$, when the mirrors are fully reflective, meaning that the Gaussian process is applied directly to the strong coherent mode, and also the output of the process can be measured with homodyne measurement. This is useful if we are estimating only the means, but completely kills method (i) since then the Gaussian process is applied to a coherent source with minimal variance, which makes it impossible to reconstruct the parameters. However, in practice the value of $T$ is limited to low values, because for larger values of $T$ there an additional decoherence and noise appears \cite{OM}, which could ruin our estimation.

\subsection{The robustness of the method}

Every result presented so far is based on the assumption of an ideal scenario, but in practice noise and damping of the matter can be present. We assumed that these imperfections are added during the Gaussian process. We assumed that the critical place is the Gaussian process. We simulate it with a coupled thermal source after the Gaussian process, which adds to the matter mode a thermal noise with variance $V_C$ and a loss described by the transmittance $T_C$ . More precisely, we assume that this is caused by the actual physical settings, so it is present independently of the unknown process on the matter mode. So in principle, we can use a calibration step to estimate the noise and the loss when the measured Gaussian process is turned off. And then we can include this estimate in our formulas to get more precise estimates of the parameters of the process.

We will use Method (ii) for numerical characterization, we saw that performs better for realistic parameters, but since it uses only the mean values, it is robust against noise. We can see that our method is robust even against loss: by using the two-step algorithm the estimation efficiency does not really depend on the actual values of the loss (Fig.\ \ref{T1}, right subfigure, dashed lines). While if we do not use any calibration (we assume that the channel is ideal, even if it is not), the same does not hold, the estimation of the displacement and the squeezing is sensitive to the losses (Fig.\ \ref{T1}, dotted lines).

\section{Conclusion}

We investigated the estimation of a Gaussian process changing a matter mode, the effect of which is measured using an ancillary light mode. We proposed an interferometric setup exploiting quantum interfaces and homodyne/heterodyne detectors and two methods for the estimation of the unknown Gaussian channel: (i) based on the covariance matrix of the measured amplitudes, (ii) based only on the means of the measured amplitudes. We compared the results to non-interferometric schemes (applying the Gaussian process to the thermal noise of the matter directly or to its mixture with a coherent light) and proved the advantage of the interferometric scheme. For the estimation of matter phase-shift alone, we found that for a large-intensity coherent light, method (ii) is sufficient for optimal estimation, while if the strength of the matter source is comparable to that of the coherent light, method (i) is required to reach the optimal estimation. This demonstrates a nontrivial optimization of the light-matter interferometry. The estimation of a general Gaussian process including non-classical squeezing was analyzed in detail using both estimation methods. We showed that a larger intensity of the coherent light used in the interferometry compensates the thermal noise of the matter. We also verified that we can obtain high-precision estimates despite the moderate coupling strength of the light-matter interfaces and access to only the light mode from the interferometer. Finally, we showed that a small loss and noise joined with the estimated Gaussian process is not critical for light-matter interferometry. In conclusion, all these results suggest that the proposed light-matter interferometry is feasible in realistic settings, and therefore it can open the way for future investigations of specific experimental platforms (e.g., in quantum optomechanics).

\section*{Funding}
Czech Science Foundation (GB14-36681G).

\section*{Acknowledgments}
L.R. acknowledges the financial support of the Faculty of Science, Palacky University.

\end{document}